\documentclass[pra,aps,amsmath,amssymb,amsfonts,twocolumn,nofootinbib,longbibliography,superscriptaddress]{revtex4-2}
\usepackage{amssymb}
\usepackage{bm,mathrsfs}
\usepackage{graphicx}
\usepackage{epsfig}
\usepackage{amsmath,bbm}
\usepackage{amsfonts,amssymb}
\usepackage{times}
\usepackage{verbatim}
\usepackage[sort&compress]{natbib}
\usepackage{amsmath}
\usepackage{xcolor}
\usepackage{bm}
\usepackage{float}

\allowdisplaybreaks[4]
\usepackage[colorlinks,breaklinks,linkcolor=blue,anchorcolor=blue,citecolor=blue,urlcolor=blue]{hyperref}
\begin{document}

\title{Parity-controlled gate in a two-dimensional neutral-atom array}

\author{F. Q. Guo}
\affiliation{Center for Quantum Sciences and School of Physics, Northeast Normal University, Changchun 130024, China}

\author{Shi-Lei Su}
\email{slsu@zzu.edu.cn}
\affiliation{School of Physics, Key Laboratory of Materials Physics of Ministry of Education, and International Laboratory for Quantum Functional Materials of Henan, Zhengzhou University, Zhengzhou 450001, China}
\affiliation{Institute of Quantum Materials and Physics, Henan Academy of Science, Henan 450046, China}

\author{Weibin Li}
\email{weibin.li@nottingham.ac.uk}
\affiliation{School of Physics and Astronomy, and Centre for the Mathematics and Theoretical Physics of Quantum Non-equilibrium Systems, The University of Nottingham, Nottingham NG7 2RD, United Kingdom}

\author{X. Q. Shao}
\email{shaoxq644@nenu.edu.cn}
\affiliation{Center for Quantum Sciences and School of Physics, Northeast Normal University, Changchun 130024, China}
\affiliation{Center for Advanced Optoelectronic Functional Materials Research, and Key Laboratory for UV Light-Emitting Materials and Technology
of Ministry of Education, Northeast Normal University, Changchun 130024, China}

\begin{abstract}

We propose a parity-controlled gate within a two-dimensional Rydberg atom array, enabling efficient discrimination between even and odd parities of virtually excited control atoms by monitoring the dynamic evolution of an auxiliary atom. This is achieved through the use of spin-exchange dipolar interactions between Rydberg states and coupling between ground states and Rydberg states. For practical applications, we explore its implementation in three-qubit repetition codes and rotated surface codes featuring $XZZX$ stabilizers, enabling single-shot readout of stabilizer measurements. Comprehensive numerical simulations are conducted to assess the feasibility of the proposed approach, taking into account potential experimental imperfections such as unwanted interactions between Rydberg states, atomic position fluctuations, laser phase noise, and Rabi amplitude noise. Our study highlights the inherent advantages of the physical mechanisms underlying parity measurement, demonstrating its reliability and practicality. These findings establish our protocol as a highly promising solution for quantum error detection and computation within Rydberg atom systems, with significant potential for future experimental realizations.

\end{abstract}

\maketitle

\section{Introduction}

Parity measurement plays a pivotal role in various quantum information processing tasks, such as entanglement generation and analysis~\cite{Hans2005Fermionic, Ionicioiu2008Generalized, Saira2014Entanglement, van2019Multipartite, Guo2020Complete}, quantum error correction (QEC)~\cite{Devitt_2013Quantum, Livingston2022Experimental, Krinner2022Realizing, poole2024architecture}, and quantum algorithms~\cite{Wolfgang2015A, Dlaska2022Quantum, Fellner2022Universal, Ender2023parityquantum}. 
Traditional parity measurement is often implemented through sequential controlled-$Z$ (CZ) or controlled-NOT (CNOT) gates~\cite{Takita2016Demonstration, Bultink2020Protecting, Acharya2023Suppressing}. Alternatively, direct construction of multiqubit parity gates provides a more efficient approach~\cite{Baptiste2018Qubit, Su2020Nondestructive, Zheng2020Robust, Dlaska2022Quantum, reagor2022hardware, christensen2023scheme, Itoko2024Three}. Compared to sequential gate methods, parity gates offer the advantage of being locally equivalent to consecutive CZ or CNOT gates that share a common control or target qubit~\cite{Itoko2024Three}. This design significantly simplifies system calibration by reducing it to a single gate or a minimal number of gates per parity check~\cite{reagor2022hardware}. Such parity gates are particularly suitable for applications in logical encoding and parity-check syndrome measurements in QEC~\cite{Nielsen2002Quantum, Terhal2015Quantum, Joschka2019Quantum, Shor1995Scheme}, as they are integral to surface codes~\cite{Fowler2012Surface, Horsman2012Surface, Andersen2020Repeated, Gertler2021Protecting, reagor2022hardware, Acharya2023Suppressing, christensen2023scheme}. These codes are implemented in two-dimensional networks of entangled physical qubits and function as stabilizer codes~\cite{gottesman1997stabilizer}. In this framework, errors are detected by evaluating the parity of four adjacent qubits, typically through measurements of stabilizer operators represented as $ZZZZ$ and $XXXX$, where $Z$ and $X$ denote Pauli operations. The eigenvalues $\pm1$ of the $N$-qubit parity operator $\bigotimes_{i}^{N}Z_{i}$ (or $\bigotimes_{i}^{N}X_{i}$) correspond to even and odd parities, respectively~\cite{christensen2023scheme}.
Therefore, parity assessment is a crucial component of quantum error detection.

Rydberg atoms are highly promising candidates for quantum computing and quantum information processing. Their remarkable characteristics, such as long coherence times, the ability to scale up, and significant electric dipole moments~\cite{Saffman2010Quantum, Saffman2016Quantum, Daniel2016An, Bernien2017Probing, Wu2021A, Shi2022Quantum, Shao2024Rydberg}, highlight their potential in a range of applications, including quantum gate implementation~\cite{Isenhower2010Demonstration, Levine2019Parallel, Shao2020Selective, Liu2020Nonadiabatic, Wu2021One, McDonnell2022Demonstration, Ma2022Universal, Fu2022High, Li2022Single, Jandura2023Optimizing, Evered2023High, Ma2023High, Heimann2023Quantum}, entanglement preparation~\cite{Saffman2009Efficient, Wilk2010Entanglement, Jau2016Entangling, Levine2018High, Graham2019Rydberg, Madjarov2020High, Shao2020Selective, Hollerith2022Realizing, Ye2023A, Shaw2024Quantum}, and quantum simulation~\cite{Daniel2016An, Bernien2017Probing, Leseleuc2019Observation, Browaeys2020Many, Lienhard2020Realization, Wu2022Manipulating, Li2022Coherent, Su2023Dipolar, Malz2023Few, Kalinowski2023Non, Tran2023Measuring, Yan2023Emergent, Zeybek2023Quantum, Kim2024Realization} using techniques such as Rydberg blockade~\cite{Jaksch2000Fast, Lukin2001Dipole}, Rydberg antiblockade~\cite{Ates2007Antiblockade}, and Rydberg dressing~\cite{Johnson2010Interactions, Henkel2010Three}. 
Significant progress is being made in the rapid development of one- and two-dimensional arrays featuring individually manipulated Rydberg atoms. These atoms are precisely controlled within optical tweezers~\cite{Daniel2016An, Bernien2017Probing, Wang2020Preparation, Ebadi2021Quantum, Scholl2021Quantum, Graham2022Multi}, providing a crucial physical platform to realize fault-tolerant quantum computing in neutral atom systems. Refs.~\cite{Crow2016Improved, Auger2017Blueprint, Cong2022Hardware, Wu2022Erasure, Scholl2023Erasure, Perlin2023Fault, Kang2023Quantum, Jandura2023Optimizing, Bluvstein2024Logical, Heussen2024Measurement, pecorari2024high, jandura2024surface} have delved into error correction in Rydberg atoms through various encoding methods, with efficient parity measurements emerging as a central focus.

Recently, a nondestructive Rydberg parity measurement, termed a ``parity meter," based on Rydberg antiblockade, has been proposed in Ref.~\cite{Su2020Nondestructive}, requiring three operational procedures. However, this method is subject to significant variations in atomic spacing, since the scheme becomes inoperative even with a small deviation of less than $1\%$ in the necessary van der Waals (vdW) interaction.
In response, Ref.~\cite{Zheng2020Robust} enhances the resilience of the Rydberg parity meter against fluctuations in interatomic distance by employing the unconventional Rydberg pumping mechanism (URP)~\cite{Li2018Unconventional}. However, this adaptation does not reduce the number of operations required to implement the scheme.
Sequential or step-by-step procedures have the capacity to enhance vulnerability to errors and decoherence, potentially resulting in a reduction in the efficiency of quantum information processing~\cite{Isenhower2010Demonstration, Shao2023highfidelity,10.1063/5.0192602}.

\begin{figure}
	\centering
	\includegraphics[width=1\linewidth]{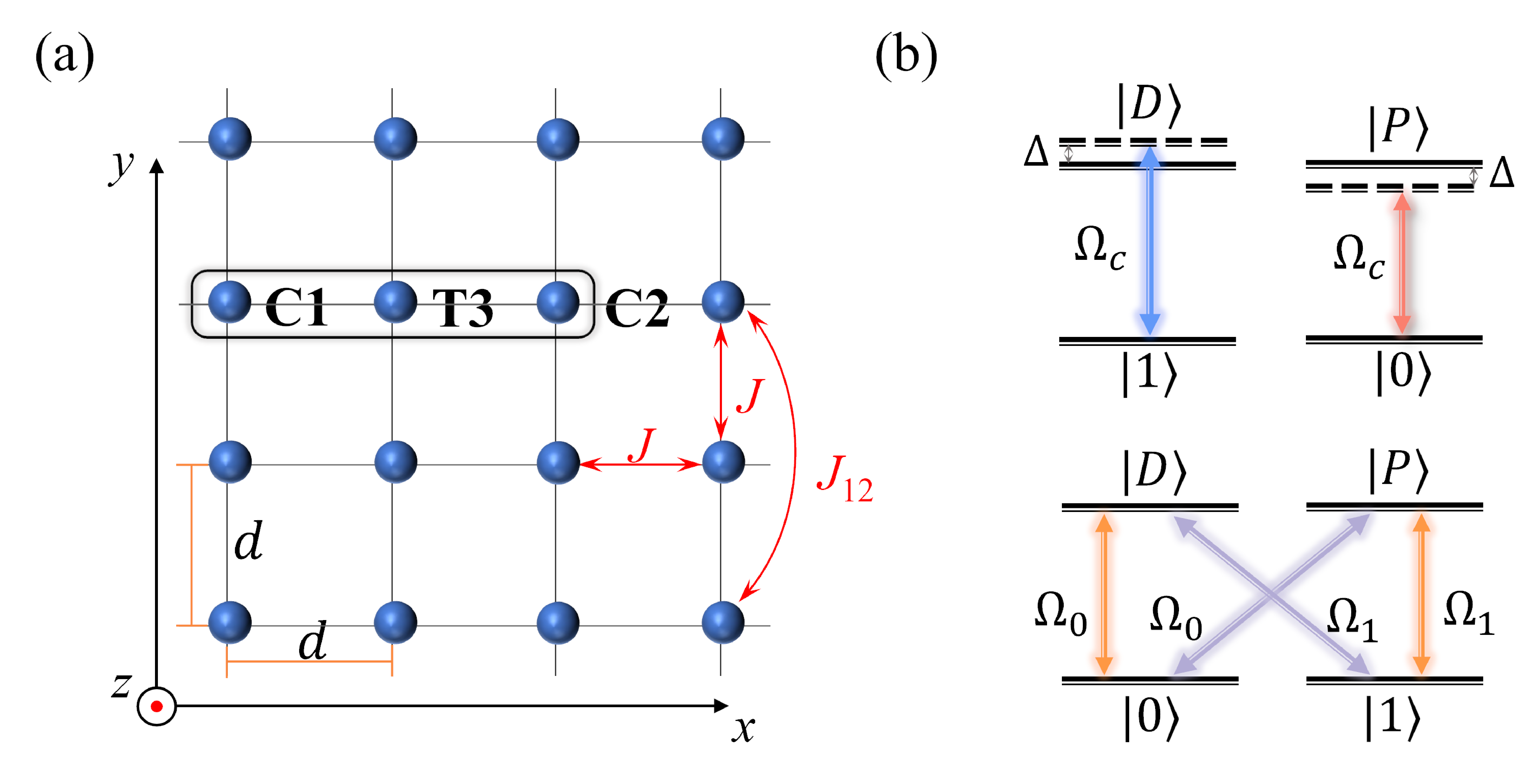}
	\caption{Atoms are arranged in a square lattice (a) with the spacing $d$ between two adjacent atoms. The dipole-dipole interaction strengths are set as $J_{13}=J_{23}=J$ between the atoms $\mathbf{C1}$, $\mathbf{C2}$, and $\mathbf{T3}$. A laser driving scheme for control atoms ($\rm \mathbf{C1},~\mathbf{C2}$, upper panel) and target atom ($\rm \mathbf{T3}$, lower panel) is shown in (b).}\label{fig1}
\end{figure}

In this work, we present a parity-controlled quantum gate tailored for neutral-atom systems. This gate leverages the dipole-dipole interaction within highly excited Rydberg states. The operation is governed by the parity of the control atoms' ground state space: different parities will either preserve the target atom's state or flip it. This design enables parity measurement of two atoms in a single operational step, eliminating the need for multiple sequential actions.
To ensure that the parity of the two control atoms remains undisturbed during the process, we apply two off-resonant lasers to induce virtual excitations to Rydberg states. Additionally, we utilize the geometric phase method~\cite{Xu2012Nonadiabatic,sjoqvist2012non} to guide the selection of a time-dependent driving field for the target atom. This approach enables a flexible choice of paths in the parameter space, allowing for efficient implementation of the desired quantum gate and offering greater versatility in system design and control optimization~\cite{Zhang20231Geometric, Liang2023Nonadiabatic, Song2024Fast}.

The remainder of this paper is structured as follows. In Sec.~\ref{sec2}, we introduce a parity-controlled gate that incorporates a geometric phase, achieved through the interaction of two control atoms and one target atom. We establish the relationships among detuning, dipole-dipole interactions, and Rabi frequencies, providing a clear representation of the effective model.
In Sec.~\ref{sec4}, we conduct a comprehensive error analysis, exploring how parity-controlled gates are affected by vdW interactions between Rydberg states, atomic position fluctuations, laser phase effects, and Rabi amplitude noise.
In Sec.~\ref{sec3}, we discuss several applications of parity-controlled gates, including a one-step Rydberg parity meter compared to previous implementations, as well as the integration of repetition and surface codes, leveraging the advantages of single-shot measurements.
Finally, in Sec.~\ref{sec5}, we summarize our findings and offer concluding remarks.

\section{PARITY-CONTROLLED ARBITRARY GATES}\label{sec2}

\subsection{Rydberg excitation engineering}\label{sec2.1}

We consider a two-dimensional square array~\cite{Daniel2016An, Ebadi2021Quantum, Scholl2021Quantum, Graham2022Multi} composed of control atoms $\mathbf{C}$ and target atoms $\mathbf{T}$, perpendicular to the quantization axis $z$ determined by a static magnetic field $B_{z}$, in which three embedded Rydberg atoms are arranged linearly. As shown in Fig.~\ref{fig1}(a), the control atoms $\mathbf{C1}$ and $\mathbf{C2}$ with the target atom $\mathbf{T3}$ are irradiated by a group of classical fields as shown in Fig.~\ref{fig1}(b). For the control atoms, the transitions $|0\rangle \leftrightarrow |D\rangle$ and $|1\rangle \leftrightarrow |P\rangle$ are driven by lasers with the same real Rabi frequency $\Omega_{c}$ but opposite detuning $\Delta$, while for the target atom, the transitions $|0\rangle \leftrightarrow |D\rangle (|P\rangle)$ and $|1\rangle \leftrightarrow |D\rangle (|P\rangle)$ are resonantly driven by lasers with complex Rabi frequencies $\Omega_{0}$ ($|\Omega_{0}|e^{i\varphi_{0}}$) and $\Omega_{1}$ ($|\Omega_{1}|e^{i\varphi_{1}}$), respectively. In the interaction picture, the corresponding Hamiltonian reads ($\hbar = 1$)
\begin{equation}\label{original}
H = \sum_{n=1}^{3} H_{n} + H_{\rm dd},
\end{equation}
where
\begin{eqnarray}
&&H_{1} = \Omega_{c}|D_{1}\rangle\langle 0_{1}|e^{-i\Delta t} + \Omega_{c}|P_{1}\rangle\langle 1_{1}|e^{i\Delta t} + {\rm H.c.},\cr\cr
&&H_{2} = \Omega_{c}|D_{2}\rangle\langle 0_{2}|e^{-i\Delta t} + \Omega_{c}|P_{2}\rangle\langle 1_{2}|e^{i\Delta t} + {\rm H.c.},\cr\cr
&&H_{3} = \Omega_{0}(|D_{3}\rangle + |P_{3}\rangle)\langle 0_{3}| + \Omega_{1}(|D_{3}\rangle + |P_{3}\rangle)\langle 1_{3}| + {\rm H.c.},\cr\cr
&&H_{\rm dd} = \sum_{i\ne j} J_{ij}|D_{i}\rangle\langle P_{i}|\otimes |P_{j}\rangle\langle D_{j}|,\nonumber
\end{eqnarray}
and $J_{ij} = C_{3}(1-3\cos^2 \Theta)/d_{ij}^{3}$ is the dipole-dipole interaction strength between atoms $i$ and $j$~\cite{Barredo2015Coherent, Leseleuc2017Optical}, with $\Theta$ the polar angle concerning the quantization axis defined by the magnetic field. In the following, we assume the atomic spacing $d_{13} = d_{23} \equiv d$, corresponding to the dipole-dipole interaction strength $J_{13} = J_{23} = J$.

\begin{figure*}
	\centering
	\includegraphics[width=1\linewidth]{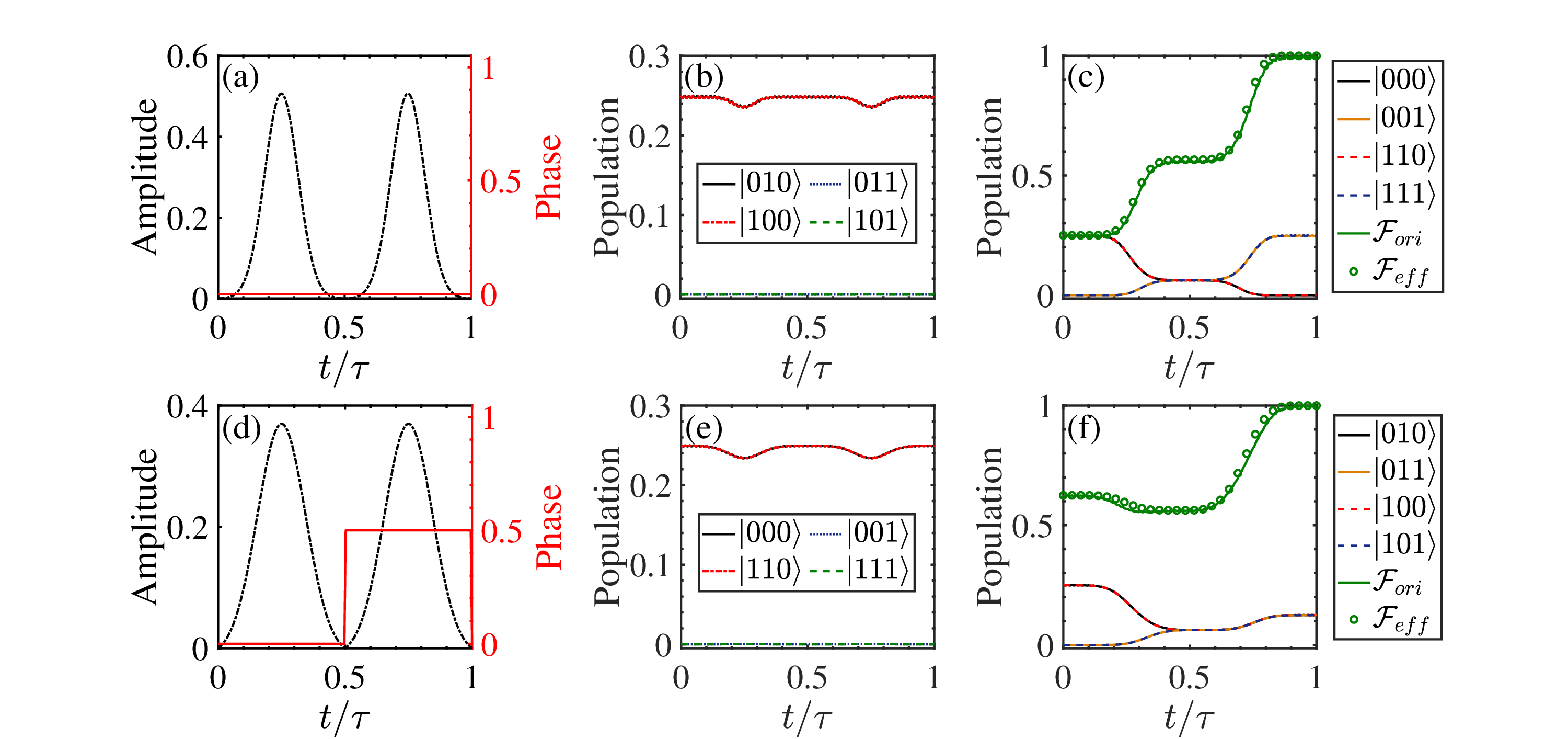}
	\caption{The upper panel and the lower panel display, separately, evolution details of gates $PE$-$X$ and $PO$-$\sqrt{X}$. (a) and (d) are pulse envelopes with unit $2\pi$~MHz and phases with unit $\pi$, where the pulses are chosen to begin and end at zero to minimize errors as much as possible. (b) and (e) show the populations of virtually excited ground states. (c) and (f) show the populations of other ground states and fidelity comparison between the original and effective Hamiltonians, where $\mathcal{F}_{ori}$ denotes the fidelity obtained from the original Hamiltonian and $\mathcal{F}_{eff}$ denotes the fidelity obtained from the effective Hamiltonian.}\label{fig2}
\end{figure*}

Under the condition $J = \Delta\gg \Omega_{c} \gg\{|\Omega_{0}|,~|\Omega_{1}|\}$, we can obtain the effective form of Eq.~(\ref{original}) as
\begin{eqnarray}\label{eq5}
\mathcal{H} &=& |0_10_2\rangle\langle 0_10_2|\otimes |D_{3}\rangle(\Omega_{0}\langle 0_{3}| + \Omega_{1}\langle 1_{3}|)\nonumber\\
&&+ |1_11_2\rangle\langle 1_11_2|\otimes |P_{3}\rangle(\Omega_{0}\langle 0_{3}| + \Omega_{1}\langle 1_{3}|) + {\rm H.c.}
\end{eqnarray}
The detailed derivation of Eq.~(\ref{eq5}) can be found in Appendix~\ref{Appendix_A}. With this Hamiltonian, it is possible to derive a three-qubit parity-controlled gate through the integration of the dressed-state process~\cite{Xu2012Nonadiabatic,sjoqvist2012non} that uses the nonadiabatic geometric way~\cite{Hong2018Implementing, Xu2018Single} to engineer the dynamical evolution. If the conditions $\Omega = \sqrt{|\Omega_{0}|^{2} + |\Omega_{1}|^{2}}$, $|\Omega_{0}|/|\Omega_{1}| = \tan{\theta/2}$, and $\varphi = \varphi_{1}-\varphi_{0}$ are met, Eq.~(\ref{eq5}) could be rewritten as
\begin{equation}\label{dress}
\mathcal{H} = \Omega e^{i\varphi_{1}}|00D\rangle\langle 00b| + \Omega e^{i\varphi_{1}}|11P\rangle\langle 11b| + {\rm H.c.},
\end{equation}
where we have used the notation $|ijk\rangle \langle ijk| = |i_{1}\rangle \langle i_{1}|\otimes |j_{2}\rangle \langle j_{2}|\otimes |k_{3}\rangle \langle k_{3}|$ for simplicity, and the ground state basis $|b\rangle = \sin{\theta/2}e^{i\varphi}|0\rangle + \cos{\theta/2}|1\rangle$ for atom $\mathbf{T3}$ and the corresponding orthogonal dark one $|d\rangle = \cos{\theta/2}|0\rangle - \sin{\theta/2}e^{-i\varphi}|1\rangle$. To build a single-loop quantum gate, we divide the evolution time $\tau$ into two halves and choose $\varphi_{1} = 0$ for $t\in[0,\tau/2)$ and $\varphi_{1} = \pi-\gamma$ for $t\in[\tau/2,\tau]$, respectively. Thus, the Hamiltonians of Eq.~(\ref{dress}) can be divided into $\mathcal{H}_{A} = \Omega(|00D\rangle\langle 00b| + |11P\rangle\langle 11b| + {\rm H.c.})$ and $\mathcal{H}_{B} = -\Omega(e^{-i\gamma}|00D\rangle\langle 00b| + e^{-i\gamma}|11P\rangle\langle 11b| + {\rm H.c.})$, leading to the effective geometric evolution operation $\mathcal{U_E}(\gamma,\theta,\varphi) = \exp(-i\int_{0}^{\tau}\mathcal{H}dt) = (|00\rangle\langle 00| + |11\rangle\langle 11|)\otimes U + (|01\rangle\langle 01| + |10\rangle\langle 10|)\otimes I$ with $I$ being the identity operator. The overall evolution time $\tau$ is derived from the condition $\int_{0}^{\tau}\Omega dt = \pi$ and the single-qubit evolution operation is written as $U(\gamma,\theta,\varphi) = |d\rangle\langle d| + e^{i\gamma}|b\rangle\langle b| = \exp(i\gamma/2)\exp(-i\gamma/2 \mathbf{n}\cdot \sigma)$ in atom 3, with $\mathbf{n}=(-\sin\theta \cos\varphi, \sin\theta \sin\varphi, \cos\theta)$ and the Pauli operator $\sigma=(X,Y,Z)$.
In addition, if we change the laser driving to $|0_{2}\rangle \stackrel{\Omega_{c}}{\longleftrightarrow} |P_{2}\rangle$ with red detuning $\Delta$ and $|1_{2}\rangle \stackrel{\Omega_{c}}{\longleftrightarrow} |D_{2}\rangle$ with blue detuning $\Delta$ on the atom $\mathbf{C2}$ while doing nothing on the atoms $\mathbf{C1}$ and $\mathbf{T3}$, we are able to get an odd parity-controlled gate $\mathcal{U_O}(\gamma,\theta,\varphi) = (|00\rangle\langle 00| + |11\rangle\langle 11|)\otimes I + (|01\rangle\langle 01| + |10\rangle\langle 10|)\otimes U$. 

\subsection{Effectiveness verification}\label{sec2B}

We choose ground states $|0\rangle \equiv |5S_{1/2}, F=2,m_{F}=0\rangle$ and $|1\rangle \equiv |5S_{1/2}, F=2,m_{F}=2\rangle$, both of which can be coupled to the Rydberg state $|D\rangle \equiv |nD_{5/2},m_{j}=5/2\rangle$ with {two-photon} excitation~\cite{Liu2021Infidelity, Fu2022High}, and coupled to $|P\rangle \equiv |n'P_{3/2},m_{j}=3/2\rangle$ with single-photon excitation~\cite{Tong2004Local, Thoumany2009Optical, Wuster2011Excitation, Glaetzle2017A, Li2019Optical, Srakaew2023A} or {three-photon} excitation~\cite{, Ryabtsev2011Doppler, Beterov2018Fast, Cheinet2020Three, Ashkarin2022Toffoli, photonics2023Two}. Similarly, excitation $|0\rangle\leftrightarrow |D\rangle$ (and $|1\rangle\leftrightarrow |P\rangle$) with blue (or red) detuning $\Delta$ for the control atoms can be executed.
\begin{figure}
	\centering
	\includegraphics[width=1\linewidth]{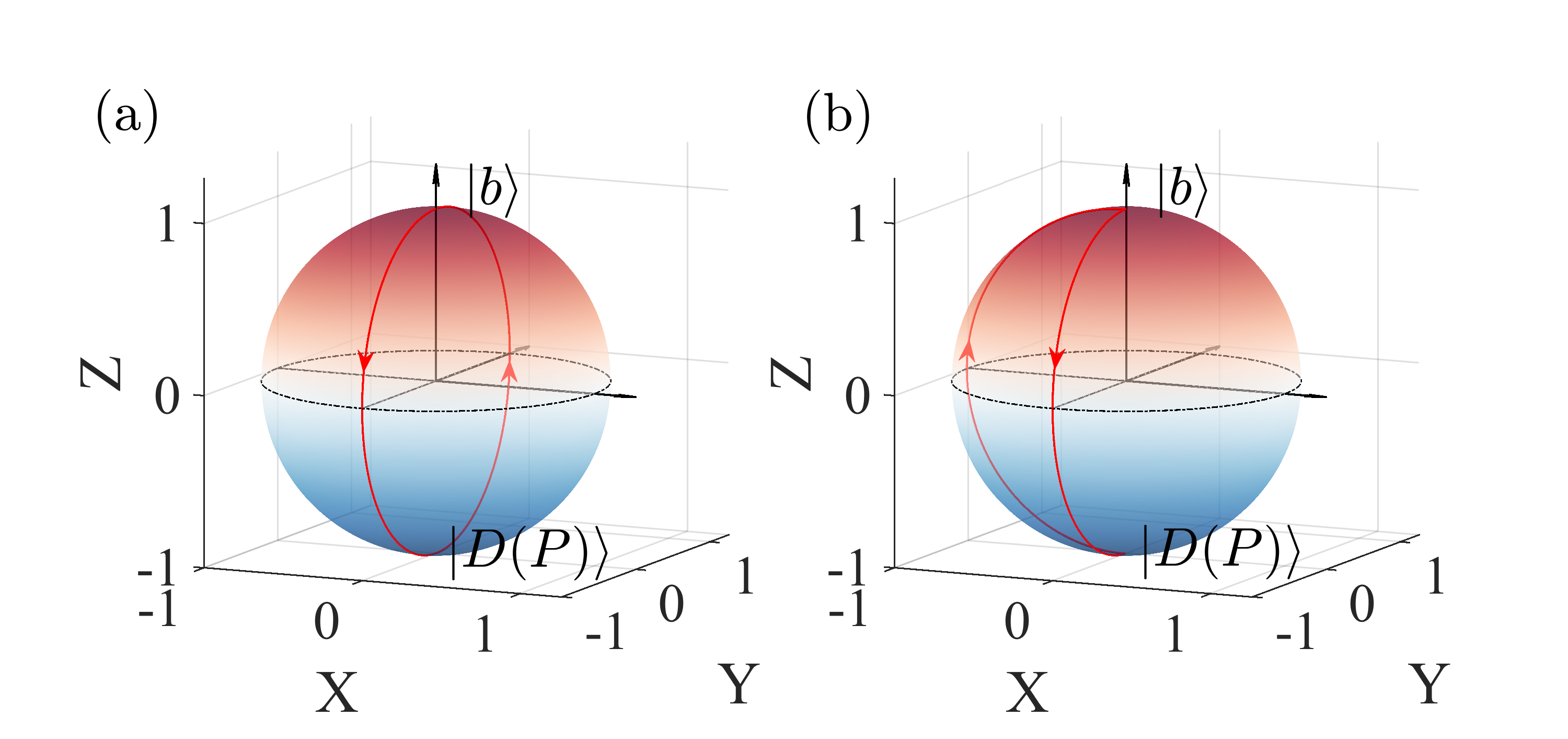}
	\caption{Diagrams showing the rotation on atom $\mathbf{T3}$ for Pauli $X$ (a) supplied by $U(\pi,\pi/2,\pi)$, and $\sqrt{X}$ (b) given by $U(\pi/2,\pi/2,\pi)$. The two rotation operations always begin and end at the $|b\rangle$ (north pole), passing through the $|D(P)\rangle$ (south pole) in the process. The evolution trajectories on the Bloch sphere are enclosed in a circle (orange-slice-shaped path), with the open angle controlled by the geometric phase.}\label{fig3}
\end{figure}

The Rabi frequency $\Omega_c$ is set to be time-independent, while $\Omega_0$ and $\Omega_1$ are time-dependent, such that each segment of $\Omega(t)$ follows a Gaussian pulse
\begin{eqnarray}
 \Omega(t) = \begin{cases} \Omega_{f}[e^{\frac{-(t-2T)^2}{2(\alpha T)^2}}-a]/(1-a), & 0 < t \leq 4T \\ \Omega_{f}[e^{\frac{-(t-6T)^2}{2(\alpha T)^2}}-b]/(1-b), & 4T < t \leq \tau \end{cases}
\end{eqnarray}
with the maximum value $\Omega_{f}$ being independent of time, $T = \tau/8$, $a = \exp\{-(2T)^2/[2(\alpha T)^2]\}$, and $b = \exp\{-(\tau/2 - 6T)^2/[2(\alpha T)^2]\}$ ensuring the amplitude is zero at the start and end~\cite{Saffman2020Symmetric}. This Gaussian time-dependent soft control can effectively mitigate the impact of off-resonant terms, even at higher driving intensities, compared to time-independent driving~\cite{PhysRevLett.121.050402}.
For the numerical simulation evaluated by Schr\"{o}dinger equation of Eq.~(\ref{original}), we initialize the system with the state $|\psi_{0}\rangle = 1/2(|0_{1}\rangle + |1_{1}\rangle) \otimes (|0_{2}\rangle + |1_{2}\rangle) \otimes |0_{3}\rangle$, and the temporal evolution of fidelity, denoted as $\mathcal{F} = |\langle \psi_0|\mathcal{U^{\dag}_{E(O)}}|\psi(t)\rangle|^2$, is measured by the population dynamics of the target state. To determine the optimal parameters $\Omega_{c}$, $\Omega_{f}$, and $\alpha$ without accounting for noise, we begin by presetting $\Omega_c = 2\pi\times 4.3~{\rm MHz}$, $J = \Delta = 20\Omega_{c}$, the evolution duration $\tau = 3~{\mu}$s, $\Omega_{f} = 2\pi\times 0.4~{\rm MHz}$, and $\alpha = 0.3$. With fixed $J,~\Delta$, and $\tau$, we then search for the optimal values of $\Omega_{c}$, $\Omega_{f}$, and $\alpha$, by minimizing the cost function $1 - \mathcal{F}$. This optimization is performed using the \texttt{minimize} function in Python or \texttt{fmincon} in MATLAB.
As a result, the optimized parameters are found to be $\Omega_{c} = 2\pi\times 3.533~{\rm MHz}$, $\Omega_{f} = 2\pi\times 0.5069~{\rm MHz}$, and $\alpha = 0.529$ for the even parity-controlled rotation $PE$-$X$ [$\equiv \mathcal{U_E}(\pi,\pi/2,\pi)$], and $\Omega_{c} = 2\pi\times 2.3455~{\rm MHz}$, $\Omega_{f} = 2\pi\times 0.3699~{\rm MHz}$, and $\alpha = 0.7584$ for the odd parity-controlled rotation $PO$-$\sqrt{X}$~[$\equiv \mathcal{U_O}(\pi/2,\pi/2,\pi)$]. In the following section, these parameters will be re-optimized to account for realistic experimental conditions. Figure~\ref{fig2} presents both pulses and populations versus time. Specifically, Figs.~\ref{fig2}(a) and~\ref{fig2}(d) depict the pulse amplitudes $\Omega(t)$ and phases $\varphi_{1}(t)$, respectively. These results correspond to $PE$-$X$ on atom $\mathbf{T3}$ with the evolution operation $U(\pi,\pi/2,\pi)$ and $PO$-$\sqrt{X}$ on the same atom with $U(\pi/2,\pi/2,\pi)$. For clarity, the rotation operations applied to the atom $\mathbf{T3}$ are illustrated in Fig.~\ref{fig3}.
Figures \ref{fig2}(b)-\ref{fig2}(c) and \ref{fig2}(e)-\ref{fig2}(f) display the population evolution for each computational basis, while the corresponding fidelities for the original and effective systems are presented in Figs.~\ref{fig2}(c) and \ref{fig2}(f). The observed changes in the consistency of the fidelity curves highlight the effectiveness of the proposed scheme. Furthermore, we find that the atoms $\rm \mathbf{C1},~\mathbf{C2}$ always stay in the ground states, for example, in the case of $PE$-$X$, the ground states $\{|010\rangle, |011\rangle, |100\rangle, |101\rangle\}$ are almost unchanged except for slight oscillations while the ground states $\{|000\rangle, |001\rangle, |110\rangle, |111\rangle\}$ only have mutual exchange, which corresponds to the effective Hamiltonian in Eq.~(\ref{eq5}).

\section{Discussion of the experimental feasibility}\label{sec4}
\begin{figure}
	\centering
	\includegraphics[width=1\linewidth]{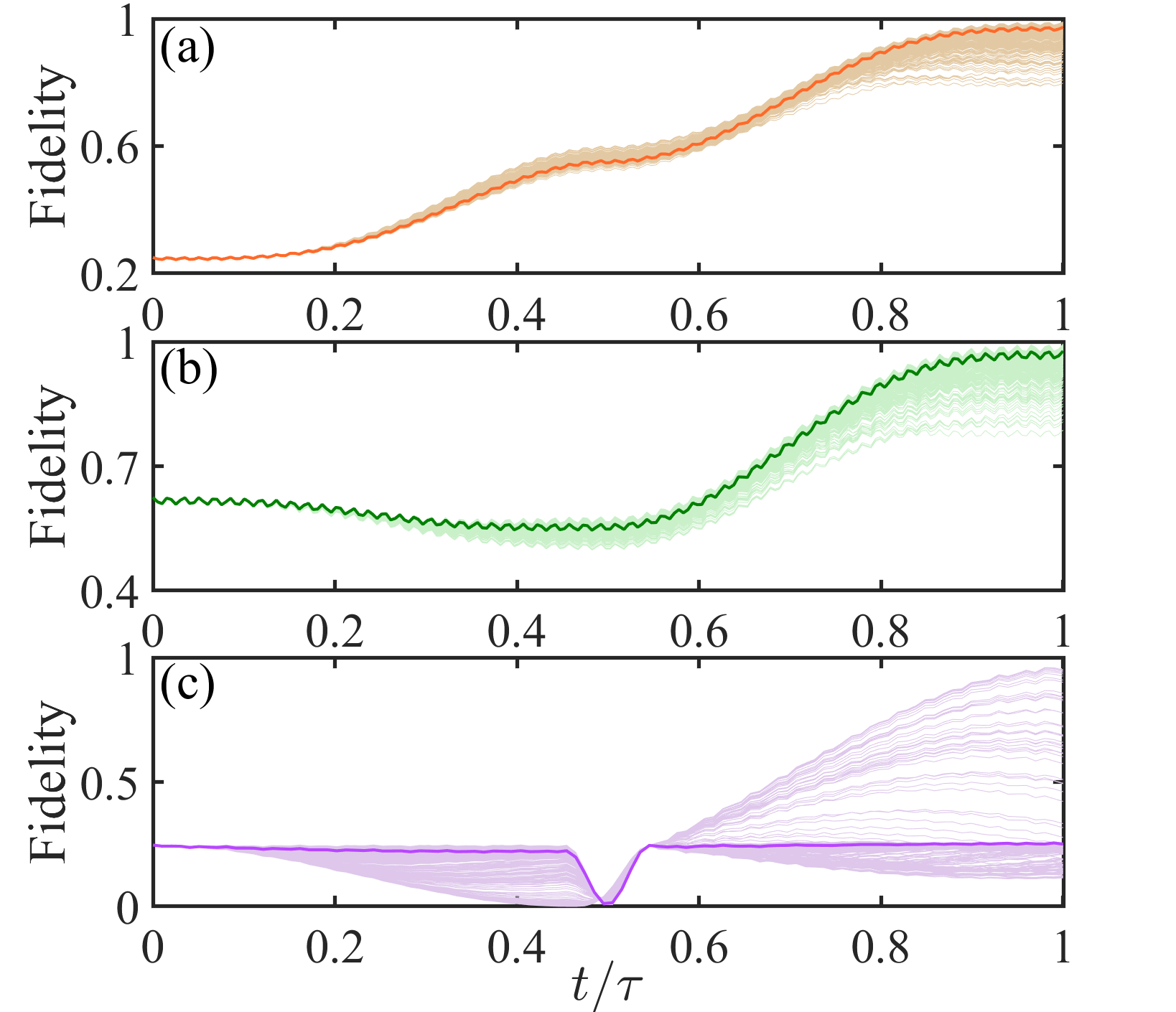}
	\caption{Fidelities against position fluctuations for gates $PE$-$X$ (a) and $PO$-$\sqrt{X}$ (b) simulated by Monte Carlo integration~\cite{Shi2017Annulled}, where each light-colored line denotes fidelity $\mathcal{F}$ subjected to different position fluctuations and a bright-colored line denotes average fidelity $\Bar{\mathcal{F}}$ with the initial state $|\psi(0)\rangle = 1/2(|00\rangle + |01\rangle + |10\rangle + |11\rangle)\otimes|0\rangle$. (c) is the fidelity in Ref.~\cite{Su2020Nondestructive} against position fluctuations. Here we set $\mathcal{F} = |\langle \psi(\tau)|\psi(t)\rangle|^2$ and initial state $|\psi(0)\rangle = 1/(2\sqrt{2})(|00\rangle + |01\rangle + |10\rangle + |11\rangle)\otimes(|0\rangle + |1\rangle)$.}\label{fig4}
\end{figure}

To facilitate the experimental implementation of our scheme, we can employ $|D\rangle = |79D_{5/2}, j = 5/2, m_{j}=5/2\rangle$ and $|P\rangle = |80P_{3/2}, j = 3/2, m_{j}=3/2\rangle$~\cite{Urban2009Observation, 
 Liu2021Infidelity, Fu2022High}, thus the corresponding $C_{3} = 2\pi\times 13.0525~{\rm GHz}~{\mu}$m$^{3}$~\cite{ROBERTSON2021107814} and atomic distance $d = 5.334~{\mu}$m under the circumstance $\Theta = \pi/2$. This determines the type and magnitude of general errors. Next, we attempt to propose experimentally feasible parameters to ensure that the parity-controlled gate is carried out without being significantly affected by random or specific errors in a realistic experimental setup. Here, we mainly discuss four kinds of noise, including atomic position fluctuations, laser phase noise, Rabi amplitude noise, and spontaneous emission from Rydberg states. 
 Additionally, we must account for vdW interactions, which naturally exist in real systems. Fortunately, they only affect the modification of the Stark shifts of atoms $\mathbf{T3}$ with $\Omega_{c}^2/(\Delta - V_{13}^{D}) + \Omega_{c}^2/(\Delta - V_{23}^{D}) -2\Omega_{c}^2/\Delta$ on $|D\rangle$ and $\Omega_{c}^2/(\Delta + V_{13}^{P}) + \Omega_{c}^2/(\Delta + V_{23}^{P}) - 2\Omega_{c}^2/\Delta$ on $|P\rangle$, with $V_{ij}^{D(P)}$ being the vdW interaction strength between Rydberg atoms $i$ and $j$ (see Appendix~\ref{Appendix_B}). It should be noted that, in our scheme, the condition 
$J=\Delta$ is reminiscent of the Rydberg antiblockade mechanism. However, since the Rydberg double-excitation states are virtually excited, the scheme offers some robustness against fluctuations in atomic spacing. Despite this, position fluctuation errors remain the primary factor influencing the performance of this scheme, especially compared to the Rydberg blockade. To address this, we will perform a new numerical optimization based on the specific experimental parameters of the optical traps. The resulting optimized parameter set will also be used to evaluate other experimental errors.

\textit{Atomic position fluctuations.} Atomic position fluctuations, which come from factors such as optical trap and atomic temperature, present a challenge during pulse sequences when the traps are turned off~\cite{Shi2021Quantum}. To model the effects of these fluctuations, we first assume three optical traps that confine atoms $\rm \mathbf{C1}$, $\rm \mathbf{C2}$, and $\rm \mathbf{T3}$, with the trap centers located at $(-d,0,0)$, $(d,0,0)$, and $(0,0,0)$, respectively. Due to the finite trap depths, the atomic positions exhibit fluctuations. We denote the perturbed positions of the atoms by $(-d + \sigma_{x1},\sigma_{y1} ,\sigma_{z1})$ for $\rm \mathbf{C1}$, 
$(d +\sigma_{x2},\sigma_{y2},\sigma_{z2})$ for $\mathbf{C2}$, and $(\sigma_{x3},\sigma_{y3},\sigma_{z3})$ for $\rm \mathbf{T3}$, where $\sigma_{xi},\sigma_{yi} ,\sigma_{zi}$ represent the deviations in positions along each axis for atom $i$.
The actual dipole-dipole interaction can be calculated by interatomic distances
\begin{eqnarray}
d_{13} &=& \{[\sigma_{x3} - (\sigma_{x1} - d)]^2 + (\sigma_{y3} - \sigma_{y1})^2 \nonumber\\&&+ (\sigma_{z3} - \sigma_{z1})^2\}^{1/2},\cr\cr
d_{12} &=& \{[(\sigma_{x2} + d) - (\sigma_{x1} - d)]^2 + (\sigma_{y2} - \sigma_{y1})^2 \nonumber\\&&+ (\sigma_{z2} - \sigma_{z1})^2\}^{1/2},\cr\cr
d_{23} &=& \{[(\sigma_{x2} + d) - \sigma_{x3}]^2 + (\sigma_{y2} - \sigma_{y3})^2 \nonumber\\&&+ (\sigma_{z2} - \sigma_{z3})^2\}^{1/2}.\nonumber
\end{eqnarray}

In Ref.~\cite{Chew2022Ultrafast}, researchers demonstrated ultrafast energy exchange between two single Rydberg atoms using an optical trap with a depth of $\rm |U_{m}| = 0.62~mK$. The corresponding trap frequencies are $(\omega_{x}, \omega_{y}, \omega_{z}) = 2\pi\times (147,117,35)$~kHz, with mean motional quanta $\bar{n}_{x,y,z}=(0.11,0.11,0.56)$, and the position fluctuation $\sigma_{x,y,z} = (22,25,60)$~nm. To simulate the effects of these fluctuations, we use Monte Carlo integration~\cite{Shi2017Annulled} for numerical simulation. The random number of each atom subject to the Gaussian distribution can be generated with two uniformly distributed random numbers $\zeta_{1}$ and $\zeta_{2}$ in the interval $[0,1]$, denoted as $\sigma_{x(y,z)}\sqrt{-2\ln{\zeta_{1}}}\cos{2\pi\zeta_{2}}$~\cite{Li2022Single}. We define the average fidelity as $\Bar{\mathcal{F}} = 1/\mathcal{N} \sum_{n=1}^{\mathcal{N}} \mathcal{F}_{n}$, where the initial state is set as $|\psi(0)\rangle = 1/2(|00\rangle + |01\rangle + |10\rangle + |11\rangle)\otimes|0\rangle$.
The state $|\psi(t)\rangle$ is derived from the Schr\"{o}dinger equation $i\frac{\partial}{\partial t}|\psi(t)\rangle = H(\bm{\sigma}^{1},\bm{\sigma}^{2},\bm{\sigma}^{3},t)|\psi(t)\rangle$, where the superscripts 1, 2, and 3 denote the atoms $\mathbf{C1},\mathbf{C2},\mathbf{T3}$, and $\bm{\sigma} \equiv (\sigma_{x},\sigma_{y},\sigma_{z})$. Combining the dipole-dipole interaction and the vdW interaction influenced by position fluctuation and the preset parameters in Sec.~\ref{sec2B}, by minimizing the cost function $1 - \mathcal{\bar{F}}$, we obtain the optimal parameters $\Omega_{c} = 2\pi\times 5$~MHz, $\Omega_{f} = 2\pi\times 0.248$~MHz, and $\alpha = 42.76$ for gate $PE$-$X$
as well as $\Omega_{c} = 2\pi\times 5$~MHz, $\Omega_{f} = 2\pi\times 0.2459$~MHz, and $\alpha = 42.7456$ for gate $PO$-$\sqrt{X}$. We find the average fidelity $\Bar{\mathcal{F}} = 0.9731$ for operations $PE$-$X$ and $\Bar{\mathcal{F}} = 0.9773$ for $PO$-$\sqrt{X}$, as illustrated in Figs.~\ref{fig4}(a) and~\ref{fig4}(b). Using identical trap parameters, we compare the results with those of Ref.~\cite{Su2020Nondestructive} under similar position fluctuations for the initial state $|\psi(0)\rangle =1/(2\sqrt{2})(|00\rangle + |01\rangle + |10\rangle + |11\rangle) \otimes (|0\rangle + |1\rangle)$. It can be observed that the fidelity depicted in Fig.~\ref{fig4}(c) decreased significantly, primarily due to disruption of the Rydberg antiblockade and phase confusion caused by the vdW interaction in the target state.

Two additional factors should be highlighted: First, the anisotropy of Rydberg interactions, including dipole-dipole and vdW interactions between states $|D\rangle$ and $|P\rangle$, is neglected due to the minimal variations in angle
$\Theta$ resulting from atomic position fluctuations. Second, reducing the atomic temperature decreases position uncertainty, thus enhancing the fidelity. In particular, recent studies have shown that Rydberg atoms can be cooled to temperatures below 5~$\mu$K~\cite{Nikolaus2021Raman, Hwang2021EIT, Zhao2023Floquet}.

\begin{figure}
	\centering
	\includegraphics[width=1\linewidth]{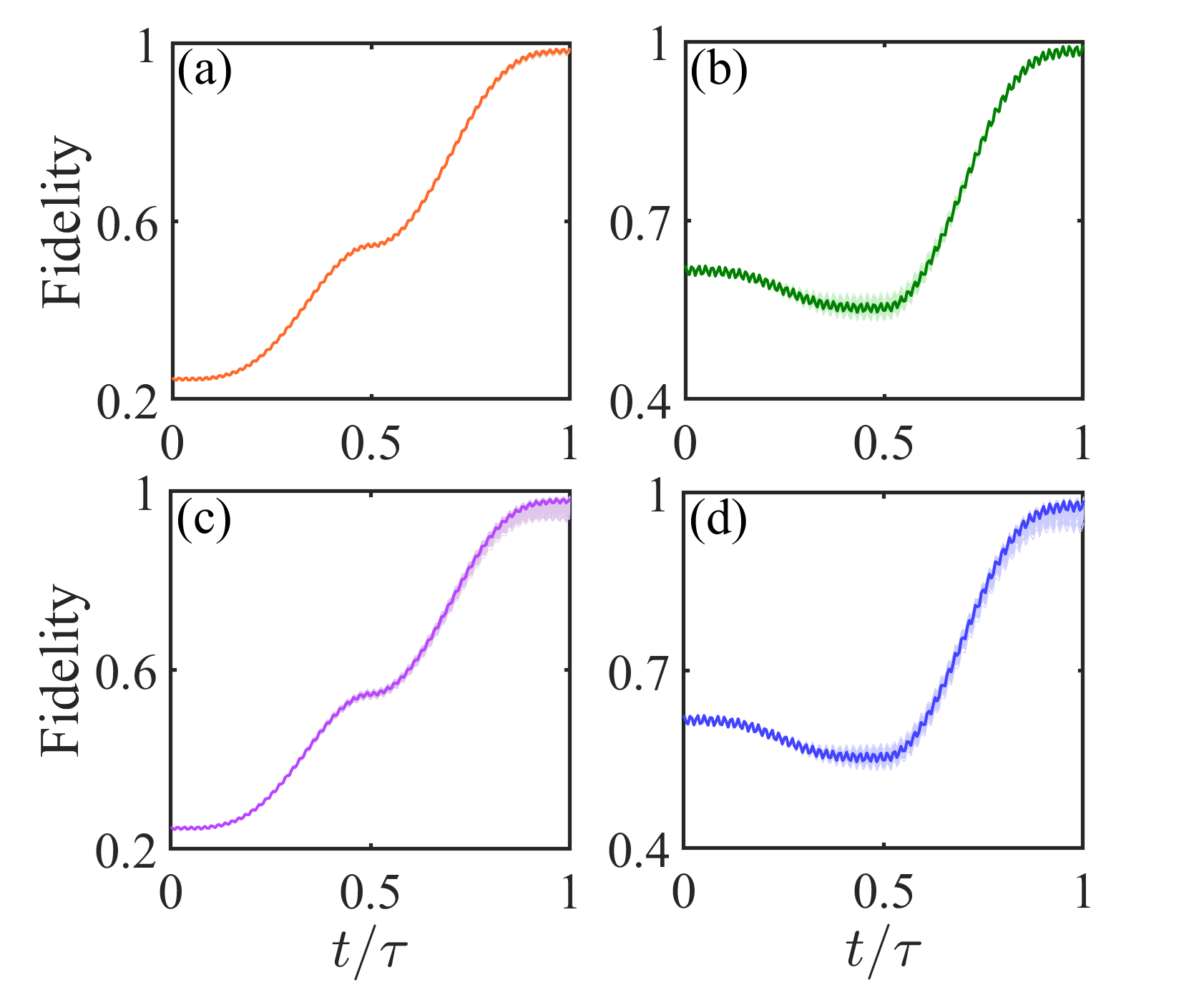}
	\caption{Fidelities against laser phase noise for gates $PE$-$X$ (a) and $PO$-$\sqrt{X}$ (b), and against Rabi amplitude noise for gates $PE$-$X$ (c) and $PO$-$\sqrt{X}$ (d) simulated by Monte Carlo integration~\cite{Shi2017Annulled}, where each light-colored line denotes fidelity $\mathcal{F}$ subjected to different noises and a bright-colored line denotes average fidelity $\Bar{\mathcal{F}}$ with the initial state $|\psi(0)\rangle = 1/2(|00\rangle + |01\rangle + |10\rangle + |11\rangle)\otimes|0\rangle$.}\label{fig5}
\end{figure}

\textit{Laser phase error.} Another important effect that contributes to the damping of oscillations is the finite-phase noise present in the excitation lasers~\cite{Sylvain2018Analysis, Levine2018High, Jo2020Rydberg}. Here we draw on the discussion in Ref.~\cite{Levine2018High} and use Monte Carlo to simulate the impact of phase noise~\cite{Jo2020Rydberg}. Each beam is power stabilized $<1\%$ by an acousto-optic modulator in Ref.~\cite{Levine2018High}; thus, we assume the laser phase noise of $\Delta\phi_{c} = \Delta\phi=0.01\pi$ (standard deviation). All beams consider different random sampling results, while two control atoms use the same random sampling results in the case $PE$-$X$. The numerical result is shown in Figs.~\ref{fig5}(a) and~\ref{fig5}(b) with fidelities of 0.9891 for gate $PE$-$X$ and 0.9934 for $PO$-$\sqrt{X}$; we find that fidelity is only slightly affected within such a fluctuation due to the precise control of laser phase.

\textit{Rabi amplitude noise.} The deviation of atoms from the center of the laser due to slight position shifts caused by the finite beam waist, or power drifts of the laser field, raises concerns regarding Rabi amplitude noise~\cite{Zhang2012Fidelity, Li2022Single, Li2023Proposal, Jandura2023Optimizing, tsai2024benchmarking}. For the sake of discussion, we assume that these inhomogeneities can be represented as $(1+\epsilon_{i})\Omega_{i}$~\cite{Jandura2023Optimizing}, where the standard deviation $\epsilon_{i}$ signifies the error rate associated with $\Omega_{i}$. Specifically, we delineate the situation into two categories: $(1+\epsilon_{c})\Omega_{c}$ and $(1+\epsilon)\Omega$, and we set $\epsilon_{c} = \epsilon = 0.008$~\cite{tsai2024benchmarking}. Similarly, all beams still consider different random sampling results, while two control atoms use the same random sampling results in the case $PE$-$X$. The numerical results are presented in Figs.~\ref{fig5}(c) and~\ref{fig5}(d) with fidelities of 0.9837 for gate $PE$-$X$ and 0.9868 for gate $PE$-$\sqrt{X}$. It can be observed that this error has a certain influence on the parity-controlled gates. However, its relatively minor impact can be attributed to advances in current experimental technology.

In addition, we also point out that the numerical results are largely unaffected by errors originating from the Rabi frequency $\Omega_{c}$ and phase $\phi_{c}$ for gates $PE$-$X$ and $PO$-$\sqrt{X}$. This robustness arises from the reliance of the scheme on the condition $\Delta\gg \Omega_{c} \gg\{|\Omega_{0}|,|\Omega_{1}|\}$, which does not impose strict numerical constraints on $\Omega_{c}$ and $\phi_{c}$. Upon applying the rotation framework $U_{rc}$ as described in Appendix~\ref{Appendix_A}, we observe that the coupling terms between ground and excited states, associated with $\Omega_{c}\exp{(i\phi_{c})}$, are discarded due to high-frequency oscillations. Consequently, the couplings driven by $\Omega_{c}$ and $\phi_{c}$ do not dominate the effective evolution of the system, making it highly resistant to errors. In contrast, errors associated with $\Omega(t)$ and $\phi$ significantly impact the fidelity of $PE$-$X$ or $PO$-$\sqrt{X}$. This is mainly because the effective Hamiltonian of Eq.~(\ref{eq5}) is closely related to the parameter $\Omega(t)$, which directly affects the performance of the gates.

\begin{figure}
	\centering
	\includegraphics[width=1.0\linewidth]{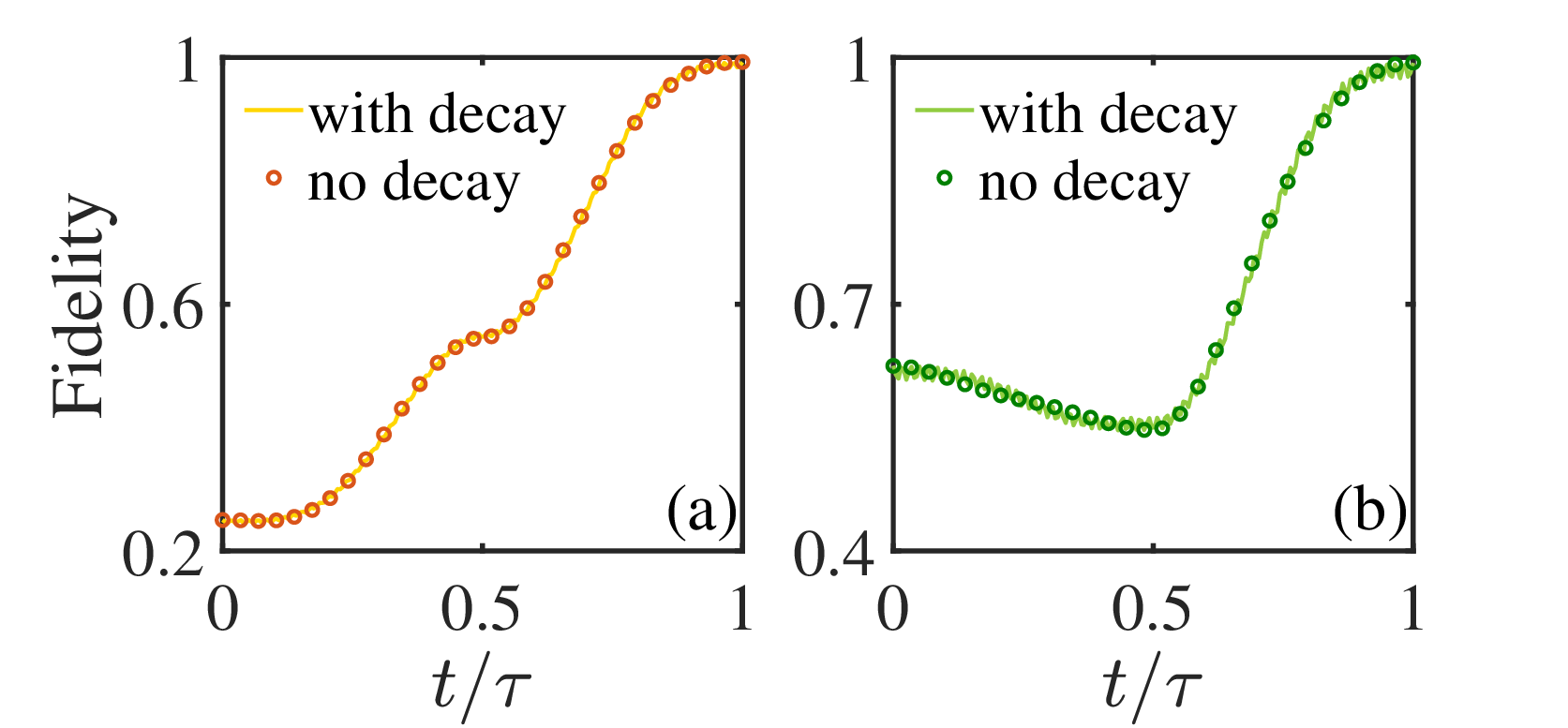}
	\caption{Fidelities against spontaneous emission for gates $PE$-$X$ (a) and $PO$-$\sqrt{X}$ (b). A solid line represents fidelity in the presence of decay, while a hollow circle represents fidelity without decay. There is no obvious difference between them.}\label{fig6}
\end{figure}

\textit{Spontaneous emission.} At the beginning, for the sake of clarity in our discussion, we introduce an uncoupled state $|m\rangle$ to encompass all ground states apart from the computational basis states. According to the Alkali-Rydberg-Calculator toolbox~\cite{ROBERTSON2021107814}, the lifetimes (in temperature $5~{\mu}$K) are $\tau_{S} = 1/\gamma_{D} = 508~{\mu}$s for the Rydberg state $|D\rangle$ and $\tau_{P} = 1/\gamma_{P} = 1140~{\mu}$s for state $|P\rangle$, and the branching ratios to dissipate the ground state manifolds are $b_{0(1)D}=b_{0(1)P}=1/8$ and $b_{mD}=b_{mP}=3/4$. The temporal evolution of the system considering atomic decay is governed by the Lindblad master equation,
\begin{eqnarray}\label{eq7}
\frac{d\rho}{dt}&=&i[\rho(t),H(t)] + \sum_{k=D,P} \mathcal{L}_{k}[\rho],
\end{eqnarray}
where
\begin{eqnarray}
\mathcal{L}_{k}[\rho] = \sum_{n = 1,2,3}\sum_{i = m,0,1} L_{ik}^{(n)}\rho L_{ik}^{(n)^{\dagger}}-\frac{1}{2}\{L_{ik}^{(n)^{\dagger}} L_{ik}^{(n)},\rho\}\nonumber
\end{eqnarray}
is the Lindblad operator describing spontaneous emission of Rydberg states $|D_{n}\rangle$ and $|P_{n}\rangle$, with $L_{ik}^{(n)} = \sqrt{b_{ik}\gamma_{k}}|i_{n}\rangle\langle k_{n}|$. The temporal evolution of fidelity $\mathcal{F}$ with Eq.~(\ref{eq7}) is changed to $\mathcal{F} = \langle \psi(\tau)|\rho(t)|\psi(\tau)\rangle$. The numerical simulation is displayed in Figs.~\ref{fig6}(a) and~\ref{fig6}(b) with fidelities of 0.9921 (decay) and 0.9927 (no error) for gate $PE$-$X$, 0.9933 (decay) and 0.9941 (no error) for gate $PE$-$X$. We find that the error caused by spontaneous emission is hardly affected regardless of whether the gate is $PE$-$X$ or $PO$-$\sqrt{X}$.

\begin{figure}
	\centering
	\includegraphics[width=1.0\linewidth]{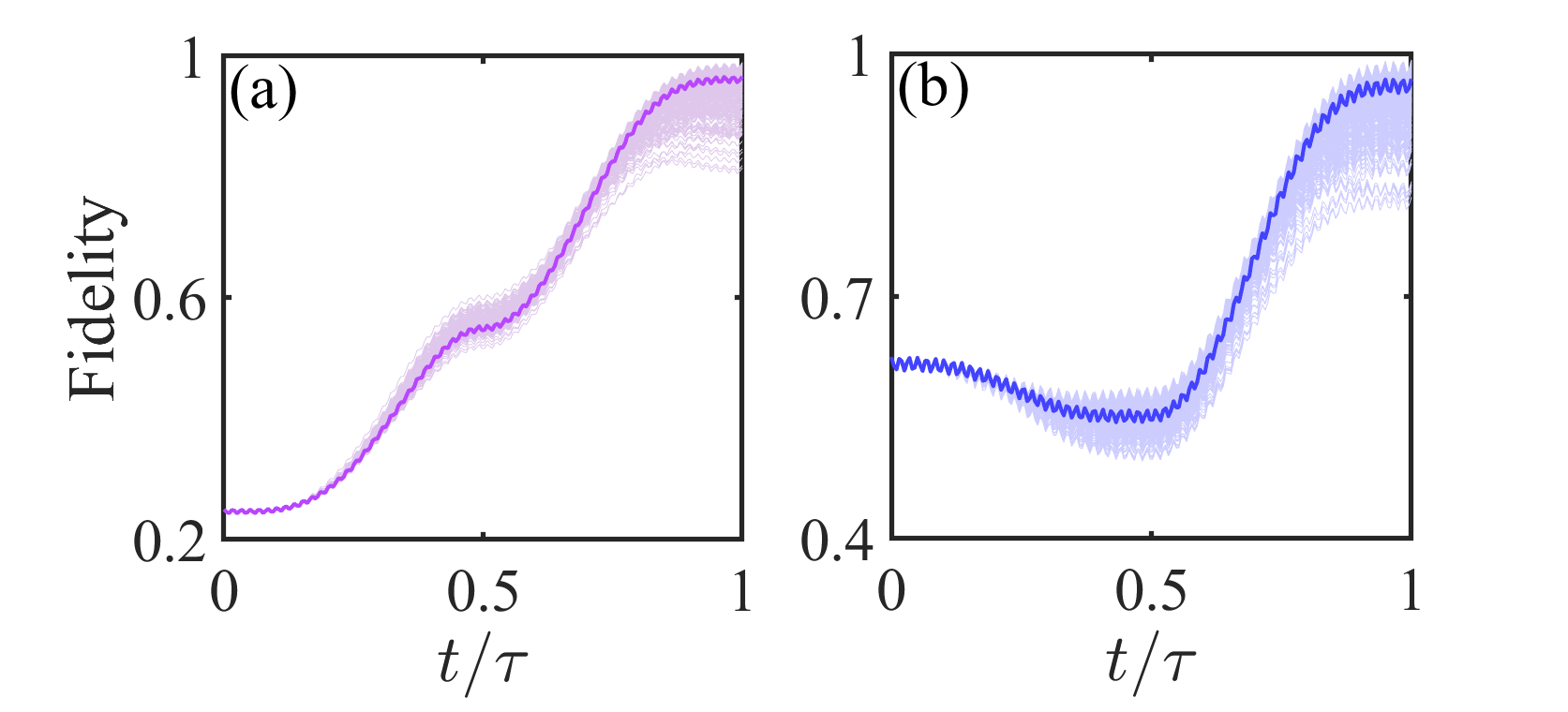}
	\caption{Fidelities after considering the vdW interactions, atomic position fluctuations, laser phase error, Rabi amplitude noise, and spontaneous emission for gates $PE$-$X$ (a) and $PO$-$\sqrt{X}$ (b). The parameters can be found in Sec.~\ref{sec4}.}\label{fig7}
\end{figure}

On balance, we also consider the influence of all errors including the vdW interactions, atomic position fluctuations, laser phase error, Rabi amplitude noise, and spontaneous emission for gates $PE$-$X$ or $PO$-$\sqrt{X}$ as shown in Fig.~\ref{fig7}. The fidelity is about 96.61$\%$ for gates $PE$-$X$ and 96.97$\%$ for gates $PO$-$\sqrt{X}$.

\section{applications of parity-controlled gate}\label{sec3}

\textit{Parity meter.} The effective evolution operation $\mathcal{U_E} = (|00\rangle\langle 00| + |11\rangle\langle 11|)\otimes U + (|01\rangle\langle 01| + |10\rangle\langle 10|)\otimes I$ is a geometric gate controlled by parity, which performs the operation $U$ on an atom $\mathbf{T3}$ when the parity of the control atoms is even, while it does nothing for an atom $\mathbf{T3}$ when odd parity. One of the most basic applications of the model is the Rydberg parity meter. Now we regard the atom $\mathbf{T3}$ as an auxiliary qubit and check the parity of the atoms $\mathbf{C1}$ and $\mathbf{C2}$. We can easily determine even or odd parity by the dynamical evolution of the auxiliary atom $\mathbf{T3}$, as mentioned above. 
The processing time of $3~{\mu}$s provided by our measurement is much less than $19.11~{\mu}$s in~\cite{Zheng2020Robust} with comparable maximum Rabi frequencies. Additionally, a comparison of our suggested method's robustness against position fluctuations in Fig.~\ref{fig4} and the parity meters covered in Ref.~\cite{Su2020Nondestructive} shows a significant improvement.

\begin{figure}
	\centering
	\includegraphics[width=1\linewidth]{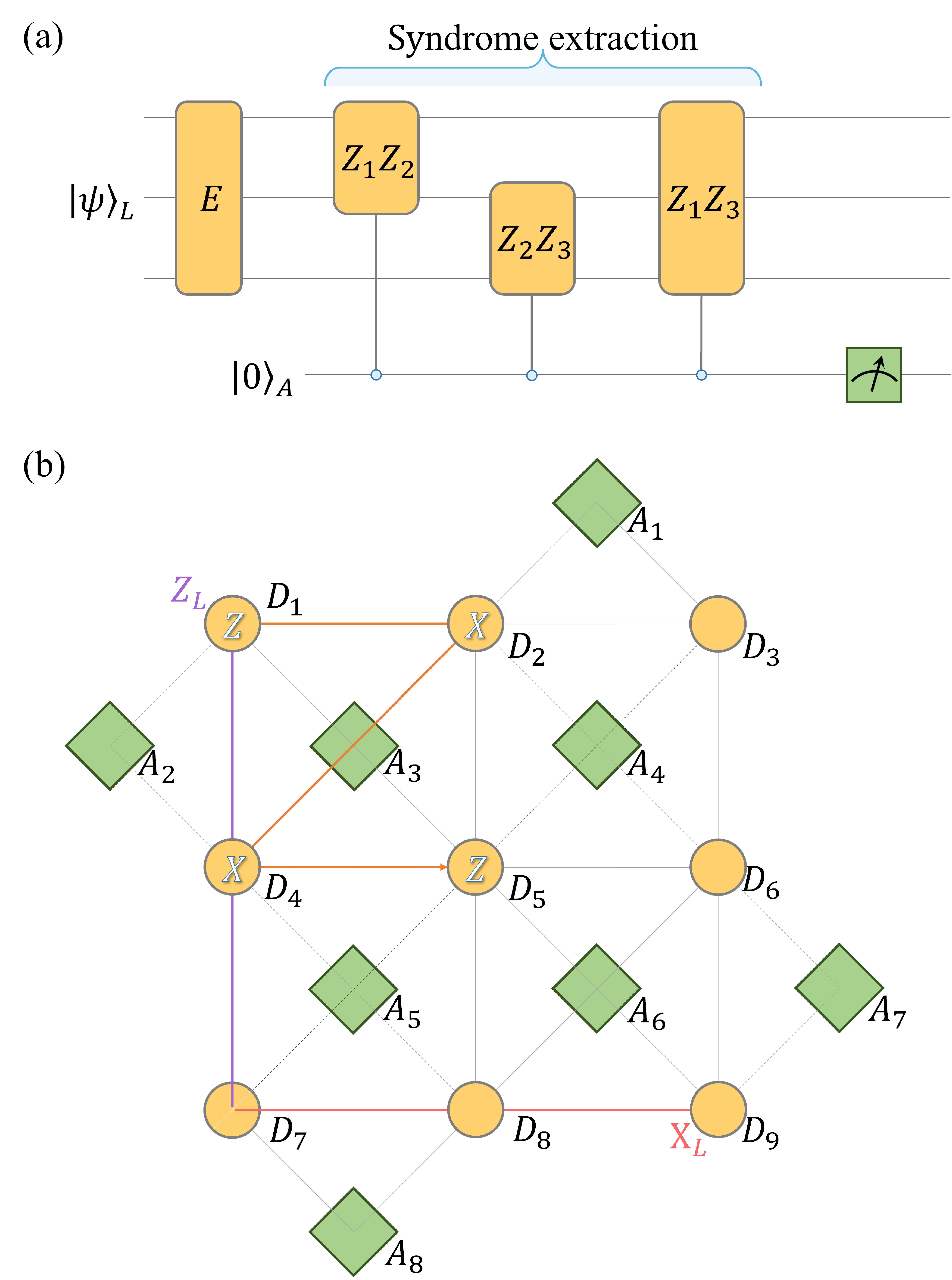}
	\caption{(a) Syndrome check of three-qubit repetition code. $E$ denotes an error that occurs in the quantum circuit. Subsequently, the three stabilizers, $Z_{1}Z_{2}$, $Z_{2}Z_{3}$, and $Z_{1}Z_{3}$ implemented by $PO$-$\sqrt{X}$ gate are applied to perform syndrome extraction. The error result is determined by the result of the single-shot measurement of the ancilla. (b) Syndrome check of $[[9,1,3]]$ rotated surface code. The yellow circle denotes a data qubit $D_{i}$ and the red square denotes an ancillary qubit $A_{i}$, {which are described by the control aotms $\mathbf{C}$ and target atoms $\mathbf{T}$, respectively}. The four-qubit stabilizer $XZZX$ can be realized in twice parity-controlled gate operations without replacement of auxiliary atoms.}\label{fig8}
\end{figure}

\textit{Three-qubit repetition code.} The logical state of this three-qubit repetition code is
\begin{equation}
|\psi\rangle_{L} = \mathcal{A}|000\rangle + \mathcal{B}|111\rangle \equiv \mathcal{A}|0\rangle_{L} + \mathcal{B}|1\rangle_{L},
\end{equation}
where the logical codewords $|0\rangle_{L}=|000\rangle$ and $|1\rangle_{L}=|111\rangle$, which can be used to correct the single-qubit flip error under such a set of basis vectors. A bit-flip error $X_{2}$ affecting only the second qubit results in
\begin{eqnarray}\label{bf_eq7}
X_{2}|\psi\rangle_{L} = \mathcal{A}|010\rangle + \mathcal{B}|101\rangle.
\end{eqnarray}
Equation~(\ref{bf_eq7}) corresponds to the error subspace $\mathcal{Q} = {\rm span}\{|100\rangle,|011\rangle,|010\rangle,|101\rangle,|001\rangle,|110\rangle\}$ transformed from the codespace $\mathcal{C} = {\rm span}\{|000\rangle,|111\rangle\}$, with $\mathcal{C}$ and $\mathcal{Q}$ being orthogonal.
Therefore, the syndrome extraction can be performed using the orthogonality of two subspaces. As shown in Fig.~\ref{fig8}(a), we use the stabilizer operator $Z_{i}Z_{j}$ to check the parity between the data qubits $i$ and $j$~\cite{Crow2016Improved}. After an error stage represented by the circuit element $E\in\{I, X_{1}, X_{2}, X_{3}\}$, we consistently observe that two stabilizer prompts exhibit errors, while one stabilizer prompt remains error-free. This strategy has been used in combination with 2-designs to improve the fault tolerance of error correction circuits~\cite{Premakumar2021designs}. In addition, its combination with our odd parity-controlled gate $PO$-$\sqrt{X}$ will yield even more unexpected results. 
When an error occurs as in Eq.~(\ref{bf_eq7}), two stabilizers will consistently show odd-parity behavior, while the remaining stabilizer will exhibit even-parity behavior. The obvious advantage of employing the $PO$-$\sqrt{X}$ scheme is that we only need one ancillary qubit and measure the ancillary qubit once throughout the process~\cite{Joschka2019Quantum}.
After threading three stabilizers $Z_{i}Z_{j}$, there are consistently two $\sqrt{X}$ operations, resulting in $\sqrt{X}\cdot\sqrt{X} = X$, so that the syndrome extraction $+1$ corresponds to the encoding space, while $-1$ corresponds to the error space. 
The repetition code corrects not only $X$-type errors, but also $Z$-type errors. To enable this, the computational basis must be transformed from $\{|0\rangle, |1\rangle\}$ to $\{|+\rangle, |-\rangle\}$, where $|+\rangle = (|0\rangle + |1\rangle)/\sqrt{2}$ and $|-\rangle = (|0\rangle - |1\rangle)/\sqrt{2}$. This transformation involves modifying the model in Fig.~\ref{fig1}(b) by introducing two additional transitions: $|S\rangle \leftrightarrow |1\rangle$ with $\Omega_{c}$ and $|P\rangle \leftrightarrow |0\rangle$ with $-\Omega_{c}$. Combining these transitions yields effective couplings $|S\rangle \leftrightarrow |+\rangle$ and $|P\rangle \leftrightarrow |-\rangle$, with a Rabi frequency of $\sqrt{2}\Omega_c$. As a result, the computational basis changes from $\{|0\rangle, |1\rangle\}$ to $\{|+\rangle, |-\rangle\}$.

\textit{Surface code.} The surface code plays an important role in QEC~\cite{Fowler2012Surface, Horsman2012Surface, Andersen2020Repeated, Chen2021Exponential, BonillaAtaides2021Performing, Marques2022Future, Krinner2022Realizing, Zhao2022Realization, Acharya2023Suppressing}. Unlike previous approaches that relied on CZ for stabilizer syndrome, we utilize alternative parity-controlled gates to achieve this task in Rydberg atoms. This, in turn, simplifies the number of gates and operational steps needed for the extraction of the stabilizer syndrome. As shown in Fig.~\ref{fig8}(b), we give the rotated $XZZX$ surface code $[[9,1,3]]$ with 9 being the total number of bits per codeword, 1 being the number of encoded bits and 3 being the code distance, which is the smallest surface code capable of detecting and correcting errors and can be represented as a single logical qubit. The symbol $D_{i}$ denotes the $i$th data qubit and $A_{i}$ denotes the ancillary one. Consequently, we have the logical $Z_{L}$ operation $Z_{D_1}Z_{D_4}Z_{D_{7}}$ and the logical $X_{L}$ operation $X_{D_7}X_{D_8}X_{D_9}$.
In this paper, our main focus lies on the error detection process. Specifically, we illustrate the application of our even parity-controlled gate $PE$-$X$ within the surface code using example stabilizers $Z_{D_1}X_{D_2}X_{D_4}Z_{D_5}$ with an auxiliary qubit $A_{3}$. The parity-controlled gate can check the parity of two atoms, thus, we need two steps to finish the extraction of four stabilizer syndromes. We can measure $Z_{D_1}Z_{D_5}$ and then $X_{D_2}X_{D_4}$ (the computational basis once again needs to be transformed from $\{|0\rangle, |1\rangle\}$ to $\{|+\rangle, |-\rangle\}$).
The parities of the dual measurements coincide with the outcome of the ultimate stabilizer measurement. Specifically, an odd count of occurrences with odd parity corresponds to $-1$, while an even count with odd parity corresponds to $+1$.
For example, considering the four-atom state $|\psi\rangle_{\rm four} = (|0+-0\rangle_{1245} + |1-+1\rangle_{1245})/\sqrt{2}$, we obtain the stabilizer $Z_{1}X_{2}X_{4}Z_{5}$ result as $-1$.
Among them, we notice that the stabilizer check process is similar to the repetition code one, which does not require the re-initialization of ancillary atoms during the two parity-controlled gates.

\section{summary}\label{sec5}

The proposed approach presents a method for implementing a parity-controlled gate using spin-exchange dipole-dipole interactions between Rydberg atoms. This quantum gate demonstrates a degree of error resilience by effectively utilizing virtual excitations into Rydberg states of control atoms and leveraging the robustness offered by geometric phases. The efficiency of our parity meter is evident in its simplified procedure and enhanced processing speed, outperforming previous Rydberg-based parity methods. It is applied to three-qubit repetition codes and rotated surface codes in the $XZZX$ architecture, demonstrating its versatility. By strategically manipulating additional atoms during parity measurements, we can obtain immediate readings of stabilizers in these codes.
Furthermore, our analytical implementation of the parity-controlled gate provides deeper insights into the physical problem, offering a clearer understanding of how atom-light interactions influence the outcome. It can also be generalized to other quantum tasks, such as one-step multipartite controlled-NOT gates and the purification of multipartite entangled states, thus expanding the toolkit for quantum information processing with Rydberg atoms. These applications will be explored further in our future research, and we look forward to the experimental realization of our approach.

\section{Acknowledgment}
The authors gratefully acknowledge the anonymous reviewers for their careful review and valuable suggestions, which have significantly improved the clarity and readability of the manuscript.
This work was supported by the National Natural Science Foundation (Grant Nos. 12174048 and 12274376), a major science and technology project of Henan Province under Grant No. 221100210400, and the Natural Science Foundation of Henan Province under Grant No. 212300410085. W.L.
 acknowledges support from the EPSRC through Grant No. EP/W015641/1 and the Going Global Partnerships Programme of the
 British Council (Contract No. IND/CONT/G/22-23/26)

\section{Author Contributions}
X.Q.S. conceptualized the study. F.Q.G. drafted the manuscript and conducted data analysis under the guidance of S.L.S., W.B.L., and X.Q.S. All authors contributed to the review and editing of the manuscript.

\appendix
\section{Derivation of effective Hamiltonian}\label{Appendix_A}

By extending the dipole-dipole interaction into three-atom direct product states, we have
\begin{equation}
    H_{\rm dd} = H_{\rm dd}^{1} + H_{\rm dd}^{2},
\end{equation}
where
\begin{eqnarray}
H_{\rm dd}^{1} &=& J(|DDP\rangle\langle PDD| + |DPP\rangle\langle PPD| + |DDP\rangle\langle DPD| \cr\cr
&&+ |PDP\rangle\langle PPD|)+ J_{12}(|DPD\rangle\langle PDD| \cr\cr
&&+ |DPP\rangle\langle PDP|) + {\rm H.c.},\cr\cr
H_{\rm dd}^{2} &=& \sum_{\Lambda=0,1}J(|D\Lambda P\rangle\langle P\Lambda D| + |\Lambda DP\rangle\langle \Lambda PD|)\cr\cr
&& + J_{12}|DP\Lambda\rangle\langle PD\Lambda| + {\rm H.c.}
\end{eqnarray}
Indeed, there exists a splitting for Rydberg energy levels due to strong dipole-dipole interaction between two Rydberg atoms. Thus, we can diagonalize $H_{\rm dd}$ according to the subsystem. First, $H_{\rm dd}^{1} = H_{\rm dd}^{1A} + H_{\rm dd}^{1B}$ includes two subspaces $\{|DPD\rangle,|PDD\rangle,|DDP\rangle\}$ and $\{|DPP\rangle,|PDP\rangle,|PPD\rangle\}$. For $H_{\rm dd}^{1A}$, we have
\begin{eqnarray}
H_{\rm dd}^{1A} &=& \mathcal{E}_1|\mathcal{E}_1\rangle\langle \mathcal{E}_1| + \mathcal{E}_2|\mathcal{E}_2\rangle\langle \mathcal{E}_2| + \mathcal{E}_3|\mathcal{E}_3\rangle\langle \mathcal{E}_3|, \cr\cr
|\mathcal{E}_1\rangle &=& -\frac{1}{\sqrt{2}}|DPD\rangle + \frac{1}{\sqrt{2}}|PDD\rangle, \cr\cr
|\mathcal{E}_2\rangle &=& \frac{\mathcal{E}_2}{\sqrt{2\mathcal{E}_2^{2}+4J^{2}}}|DPD\rangle + \frac{\mathcal{E}_2}{\sqrt{2\mathcal{E}_2^{2}+4J^{2}}}|PDD\rangle \cr\cr
&&+ \frac{2J}{\sqrt{2\mathcal{E}_2^{2}+4J^{2}}}|DDP\rangle, \cr\cr
|\mathcal{E}_3\rangle &=& \frac{\mathcal{E}_3}{\sqrt{2\mathcal{E}_3^{2}+4J^{2}}}|DPD\rangle + \frac{\mathcal{E}_3}{\sqrt{2\mathcal{E}_3^{2}+4J^{2}}}|PDD\rangle \cr\cr
&&+ \frac{2J}{\sqrt{2\mathcal{E}_3^{2}+4J^{2}}}|DDP\rangle, \nonumber
\end{eqnarray}
with eigenvalues $\mathcal{E}_1=-J_{12},~\mathcal{E}_2 = \frac{1}{2}(J_{12}-\sqrt{8J^2 + J_{12}^2})$, and $\mathcal{E}_3 = \frac{1}{2}(J_{12}+\sqrt{8J^2 + J_{12}^2})$. Similarly, for $H_{\rm dd}^{1B}$ we have the same eigenvalues and the same eigenstates form with different bases $\{|DPP\rangle,|PDP\rangle,|PPD\rangle\}$.

Second, $H_{\rm dd}^{2} = H_{\rm dd}^{2A} + H_{\rm dd}^{2B} + H_{\rm dd}^{2C}$ includes three subspaces, $\{|D\Lambda P\rangle, |P\Lambda D\rangle\}$, $\{|\Lambda DP\rangle, |\Lambda PD\rangle\}$, and $\{|DP\Lambda\rangle, |PD\Lambda\rangle\}$. For $H_{\rm dd}^{2A}$, we have

\begin{eqnarray}
H_{\rm dd}^{2A} &=& \mathcal{G}_1|\mathcal{G}_1\rangle\langle \mathcal{G}_1| + \mathcal{G}_2|\mathcal{G}_2\rangle\langle \mathcal{G}_2|, \cr\cr
|\mathcal{G}_1\rangle &=& -\frac{1}{\sqrt{2}}|D\Lambda P\rangle + \frac{1}{\sqrt{2}}|P\Lambda D\rangle, \cr\cr
|\mathcal{G}_2\rangle &=& \frac{1}{\sqrt{2}}|D\Lambda P\rangle + \frac{1}{\sqrt{2}}|P\Lambda D\rangle,\nonumber
\end{eqnarray}
with eigenvalues $\mathcal{G}_1 = -J,~\mathcal{G}_2 = J$. Similarly, for $H_{\rm dd}^{2B}$ we have the same eigenvalues and the same eigenstates form with different bases $\{|\Lambda DP\rangle, |\Lambda PD\rangle\}$. While for $H_{\rm dd}^{2C}$ we have

\begin{eqnarray}
H_{\rm dd}^{2C} &=& \mathcal{O}_1|\mathcal{O}_1\rangle\langle \mathcal{O}_1| + \mathcal{O}_2|\mathcal{O}_2\rangle\langle \mathcal{O}_2|, \cr\cr
|\mathcal{O}_1\rangle &=& -\frac{1}{\sqrt{2}}|DP\Lambda\rangle + \frac{1}{\sqrt{2}}|PD\Lambda\rangle, \cr\cr
|\mathcal{O}_2\rangle &=& \frac{1}{\sqrt{2}}|DP\Lambda\rangle + \frac{1}{\sqrt{2}}|PD\Lambda\rangle,\nonumber
\end{eqnarray}
with eigenvalues $\mathcal{O}_1 = -J_{12}, \mathcal{O}_2 = J_{12}$.

For briefness, we use the notion $E_{\nu}^{\mu}$ to denote the ${\mu}$th basis of $\nu$ subspace, where ${\mu}\in\{1,2,3\}$ and $\nu\in\{DPD,DPP,D\Lambda P,\Lambda DP,DP\Lambda$\}. Assuming the condition $J = \Delta\gg \{|\Omega_{0}|,|\Omega_{1}|,\Omega_{c}\}$, then a rotation $U_{rd}=\exp(-iH_{\rm dd}t)$ is performed. The total Hamiltonian $H$ after eliminating the high-frequency oscillating terms can be written as

\begin{eqnarray}\label{eq3}
H = H_{00} + H_{01} + H_{10} + H_{11},
\end{eqnarray}
where
\begin{eqnarray}
H_{00} &=& \Omega_0|00D\rangle\langle 000|+\Omega_1|00D\rangle\langle 001| + \Omega_0|00P\rangle\langle 000|\cr\cr
&&+ \Omega_1|00P\rangle\langle 001| + \frac{\Omega_{c}}{\sqrt{2}}|E^{2}_{D0P}\rangle\langle 00P| \cr\cr
&&+ \frac{\Omega_{c}}{\sqrt{2}}|E^{2}_{0DP}\rangle\langle 00P| + {\rm H.c.}, \cr\cr
H_{01} &=& \Omega_0|01D\rangle\langle 010| + \Omega_1|01D\rangle\langle 011|\cr\cr
&&+ \frac{\Omega_{c}}{\sqrt{2}}|E_{0DP}^{1}\rangle\langle 01D| + \Omega_0|01P\rangle\langle 010|\cr\cr 
&&+ \Omega_1|01P\rangle\langle 011| + \frac{\Omega_{c}}{\sqrt{2}}|E^{2}_{D1P}\rangle\langle 01P| + {\rm H.c.}, \cr\cr
H_{10} &=& \Omega_0|10D\rangle\langle 100| + \Omega_1|10D\rangle\langle 101|\cr\cr 
&&+ \frac{\Omega_{c}}{\sqrt{2}}|E_{D0P}^{1}\rangle\langle 10D|
+ \Omega_0|10P\rangle\langle 100|\cr\cr 
&&+ \Omega_1|10P\rangle\langle 101|
+ \frac{\Omega_{c}}{\sqrt{2}}|E^{2}_{1DP}\rangle\langle 10P| + {\rm H.c.}, \cr\cr
H_{11} &=& \Omega_0|11D\rangle\langle 110| + \Omega_1|11D\rangle\langle 111|\cr\cr 
&&+ \frac{\Omega_{c}}{\sqrt{2}}|E_{D1P}^{1}\rangle\langle 11D| + \frac{\Omega_{c}}{\sqrt{2}}|E_{1DP}^{1}\rangle\langle 11D|\cr\cr &&+ \Omega_0|11P\rangle\langle 110| + \Omega_1|11P\rangle\langle 111| + {\rm H.c.}
\end{eqnarray}

The Hamiltonian can be further reduced by the condition $\Omega_{c}\gg\{|\Omega_0|,|\Omega_1|\}$. For example, the strong transition term $\Omega_{c}/\sqrt{2}|E^{1}_{D0P}\rangle\langle 00P| + \Omega_{c}/\sqrt{2}|E^{1}_{0DP}\rangle\langle 00P| + {\rm H.c.}$ with $\Omega_{c}$ is further diagonalized as $\lambda_1|\lambda_1\rangle\langle \lambda_1| + \lambda_2|\lambda_2\rangle\langle \lambda_2|$ with the eigenstates $|\lambda_{1}\rangle = (|E^{1}_{D0P}\rangle - \sqrt{2}|00P\rangle + |E^{1}_{0DP}\rangle)/2$, $|\lambda_{2}\rangle = (|E^{1}_{D0P}\rangle + \sqrt{2}|00P\rangle + |E^{1}_{0DP}\rangle)/2, |\lambda_{0}\rangle = (-|E^{1}_{D0P}\rangle + |E^{1}_{0DP}\rangle)/\sqrt{2}$, and the corresponding eigenvalues $\lambda_{1} = -\sqrt{2}\Omega_c,~\lambda_{2} = \sqrt{2}\Omega_c,~\lambda_{0} = 0$. Thus, the terms $\Omega_{0}|00P\rangle\langle 000| + \Omega_{1}|00P\rangle\langle 001| + {\rm H.c.}$ are transformed into $[-1/\sqrt{2}\exp(-i\sqrt{2}\Omega_{c} t)|\lambda_{1}\rangle + 1/\sqrt{2}\exp(i\sqrt{2}\Omega_{c} t)|\lambda_{2}\rangle](\Omega_{0}\langle 000| + \Omega_{1}\langle 001|) + {\rm H.c.}$ after using the rotation frame $U_{rc} = \exp[-i(\lambda_1|\lambda_1\rangle\langle \lambda_1| + \lambda_2|\lambda_2\rangle\langle \lambda_2| + \lambda_0|\lambda_0\rangle\langle \lambda_0|)t]$ and then discarded due to high-frequency oscillation. Similar processing occurs on $H_{01}, H_{10}$, and $H_{11}$. As a result, the final Hamiltonian is as follows:
\begin{eqnarray}
H_{\rm eff} &=& |00\rangle\langle 00|\otimes (\Omega_{0}|D\rangle\langle 0| + \Omega_{1}|D\rangle\langle 1|)\cr\cr
&&+ |11\rangle\langle 11|\otimes (\Omega_{0}|P\rangle\langle 0| + \Omega_{1}|P\rangle\langle 1|)+ {\rm H.c.}
\end{eqnarray}

Several techniques can be used to mitigate Stark shifts, including phase adjustments~\cite{Veps2018Optimal, Veps2019Superadiabatic}, the use of auxiliary detuned transitions~\cite{Su2016One, Zhao2017Robust}, and the optimal determination of detunings~\cite{Wang2020Generation, Han2020Multi}. In the present case, additional lasers have been utilized to counteract the Stark shifts $\Omega_{c}^2/ \Delta$ that affect the control atoms $\mathbf{C1}$ and $\mathbf{C2}$.

\section{Influence of vdW interaction}\label{Appendix_B}

The premise of our approach is that it relies exclusively on dipole-dipole interactions, rendering the vdW interaction of the relevant Rydberg states at atomic distance $d$~\cite{Fu2022High} undesirable. In principle, this type of interaction can be ignored by expanding the interatomic distance, as done in Refs.~\cite{Leseleuc2017Optical, Barredo2015Coherent}, which employ the distance between two atoms over 20 ${\mu}$m such that the vdW interaction remains on the order of kilohertz. Although our scheme cannot eliminate the vdW interaction by similar means, fortunately, we observe that it primarily affects certain unnecessary detunings that can be regarded as high-frequency oscillating terms in Eq.~(\ref{eq3}). For clarity, an analytical discussion is provided below.

Taking into account the potential Rydberg interactions, the Hamiltonian of the system reads
\begin{eqnarray}
    H = \sum_{n=1}^{3} H_{n} + H_{\rm dd} + H_{\rm vdw},
\end{eqnarray}
where $H_{\rm vdw} = \sum_{i<j}V_{ij}^{D}|D_{i}D_{j}\rangle\langle D_{i}D_{j}| + V_{ij}^{P}|P_{i}P_{j}\rangle\langle P_{i}P_{j}|$. Here, $V_{ij}^{D, P} = - C_{6}^{D, P}(\Theta,\Phi)/d_{ij}^{6}$ denotes the vdW interaction strength with angle $(\Theta, \Phi)$ on a spherical basis. Specifically, $C_{6}^{D}(\pi/2,0) = 2\pi\times 1542.60~{\rm GHz}~{\mu}$m$^6$ for the Rydberg state $|79D_{5/2},j=5/2,m_{j} = 5/2\rangle$ and $C_{6}^{P}(\pi/2,0) = 2\pi\times 5486.82~{\rm GHz}~{\mu}$m$^6$ for $|80P_{3/2},j=3/2,m_{j} = 3/2\rangle$, resulting in $V_{13}^{D} = V_{23}^{D} = - 2\pi\times 66.978~{\rm MHz}$, $V_{12}^{D} = - 2\pi\times 1.0465~{\rm MHz}$, $V_{13}^{P} = V_{23}^{P} = - 2\pi\times 238.23~{\rm MHz}$ and $V_{12}^{P} = - 2\pi\times 3.7223~{\rm MHz}$. It is important to note that when considering these unwanted interactions, a reexamination of the Stark shifts in states $|00D\rangle$ and $|11P\rangle$ is necessary. For Hamiltonian components $H_{00}$ and $H_{11}$ in Appendix~\ref{Appendix_A}, these interactions alter the detunings of states $|00D\rangle$ and $|11P\rangle$, thereby impacting the previous scheme to eliminate Stark shifts on atoms $\mathbf{C1}$ and $\mathbf{C2}$. Regarding components $H_{01}$ and $H_{10}$ in Appendix~\ref{Appendix_A}, there are always resonant channels that suppress the evolution between ground states, such as $\Omega_{c}/\sqrt{2}|E_{0DP}^{1}\rangle\langle 01D| + {\rm H.c.}$ and $\Omega_{c}/\sqrt{2}|E_{D0P}^{1}\rangle\langle 10D| + {\rm H.c.}$, rendering the vdW term $H_{\rm vdw}$ inconsequential to the evolution of the system.
In the experiment, the Stark shifts $\Omega_{c}^2/(\Delta - V_{13}^{D}) + \Omega_{c}^2/(\Delta - V_{23}^{D}) -2\Omega_{c}^2/\Delta$ on the Rydberg states $|D\rangle$ and $\Omega_{c}^2/(\Delta + V_{13}^{P}) + \Omega_{c}^2/(\Delta + V_{23}^{P}) - 2\Omega_{c}^2/\Delta$ on $|P\rangle$ of the atom $\mathbf{T3}$ can be removed by employing two additional laser beams.

\bibliography{manuscript.bbl}

\end{document}